\documentclass[prd,tightenlines,showpacs,letterpaper,preprintnumbers,nofootinbib,floatfix]{revtex4}

\usepackage{amsmath,amssymb,amsfonts,graphicx,pifont,dcolumn}
\usepackage{epstopdf}
\usepackage{color}
\usepackage{cancel}
\usepackage{ulem}
\RequirePackage{slashed}

\begin{document}

\title{Primordial monopolium as dark matter}

\author{ Vicente Vento }

\affiliation{Departamento de F\'{\i}sica Te\'orica-IFIC, Universidad de Valencia- CSIC, 46100 Burjassot (Valencia), Spain.}

\date{\today}

\begin{abstract}

The existence of monopoles is a characteristic signature of Kaluza-Klein multidimensional theories. The topology of these solutions is extremely interesting. The existence of a dipole solution, which we have associated to a monople-anti-monopole bound state, is the leitmotiv of this investigation. The dipole in its lowest energy state, which we here call also monopolium, is electromagnetically inert in free space interacting only gravitationally.  Monopolium when interacting with time dependent magnetic fields  acquires a time dependent induced magnetic moment and radiates. We have analyzed the most favorable astrophysical scenario for radiative monopolium and found that the amount of radiation is so small that is not detectable by conventional equipments. These findings suggest that Kaluza-Klein monopolium, if existent, would be a candidate for a primordial dark matter constituent.

\end{abstract}


\maketitle

\section{introduction} \label{intro}

The nature of dark matter is one of the most relevant problems of physics today. Cosmological experiments have determined its abundance \cite{Aghanim:2018eyx}. From observational and numerical simulations we have some knowledge about its distribution in galactic halos \cite{Wechsler:2018pic}. Still the nature of dark matter remains unknown. Since the favorite explanations in the past, the weakly interacting particles, have eluded detection it is worthwhile to consider other possibilities.

Monopole  physics took a dramatic turn when 't Hooft \cite{'tHooft:1974qc} and Polyakov \cite{Polyakov:1974ek} independently discovered that the SO(3) Georgi-Glashow model \cite{Georgi:1972cj} inevitably contains monopole solutions. These topological monopoles appear naturally  in Grand Unified Theories and are impossible to create in present day particle colliders because of their huge GUT mass  \cite{Preskill:1984gd,Shnir:2005xx} and they are very rear in the universe since inflation reduces their density dramatically \cite{Guth:1980zm,Linde:1981mu}.  't Hooft-Polyakov monopoles cannot built up dark matter because of their incredibly strong interactions \cite{Dirac:1948um}. 

 It has been a long time since Kaluza and Klein first investigated the idea of unification of gravitation and gauge theories by exploring the possibility that the number of spacetime dimensions is greater than four \cite{Kaluza:1984ws,Klein:1926tv}. The idea behind Kaluza-Klein (KK) theories is that the world has more than three spatial dimensions and some of them are curled up to form a circle so small as to be unobservable. KK theories have been the subject of revived interest in recent years since many standard model extensions in extra dimensions yield KK theories. KK theories contain a very rich topological structure, which includes very heavy monopoles whose mass is of the order of the Planck mass  \cite{Gross:1983hb,Sorkin:1983ns,Newman:1963yy,Hawking:1976jb,Gibbons:1979xm}. But most important for the present study, they also contain other soliton solutions in different topological sectors. In particular,  the dipole, which has the quantum numbers of a monopole-anti-monopole bound state. This state in conventional gauge theories was called monopolium \cite{Hill:1982iq,Vento:2007vy,Epele:2007ic}, and we kept the name here. In conventional gauge theories monopolium has vacuum quantum numbers and annihilates into more elementary constituents. However, in KK theories monopolium does not belong to the topological sector of the vacuum and therefore it is classically stable  \cite{Gross:1983hb,Kerr:1963ud,Gibbons:1976ue}. That solitons of KK theories could supply dark matter has been briefly discussed in the past \cite{Overduin:1998pn}. The aim of this paper is to show that monopolium, as it appears in KK theories, if it survives inflation in sufficient numbers, could be a candidate for a primordial constituent of dark matter, and thus its study would open a window into the physics of more than four dimensions. 

 The presentation is based on the scenario discussed by Gross, Perry and Sorkin (GPS) \cite{Gross:1983hb,Sorkin:1983ns}, which serves as a toy model, but our results are based on  properties which arise in realistic scenarios. Let us describe them precisely. A topology where the dipole does not have vacuum quantum numbers~\cite{Newman:1963yy,Hawking:1976jb,Gibbons:1979xm,Kerr:1963ud,Gibbons:1976ue}. The dipole structure, described in terms of some {\it parameter} $d$ (distance between center of the poles), approaches the monopole-anti-monopole structure as  $d$ becomes large. This also occurs in topological schemes with different topological sectors like in the Skyrme Model \cite{Zahed:1986qz}. The dipole can thus be interpreted as a bound state with a mass smaller than twice the monopole mass. The parameter $d$ describes the magnetic moment of the dipole. The ground state (smallest mass) dipole, i.e. when $d\rightarrow 0$, has vanishingly small dipole moment and therefore it is electromagnetically neutral and thus has only gravitational interaction. The GPS model has all these ingredients and therefore we are going to use it as a toy model to develop the ideas and get orders of magnitude for observables. However, the GPS model has a peculiar way of realizing the equivalence principle, which we will discuss in some detail, and we will show one mechanism for avoiding it.
    
 In section~\ref{kkm} we are going to recall some aspects of the topology of the Kaluza-Klein theory in particular the existence of monopoles and dipoles. In section~\ref{dm} we analyze if the dipole survives  the  inflationary process.   In section~\ref{signals} we look for possible electromagnetic observables of monopolium and in section~\ref{astro} we discuss possible experimental observations. Their size allows us to propose monopolium as a dark matter component. We end by some concluding remarks. 

\section{Kaluza-Klein monopolium} \label{kkm}

Kaluza-Klein theories \cite{Kaluza:1984ws,Klein:1926tv} contain topological solitons which can be identified as magnetic monopoles satisfying the Dirac quantization condition, $e g = \frac{1}{2} $ \cite{Dirac:1948um}.  
We next recall the structure of the KK monopoles  by whose features can be extended to  models with more compact dimensions \cite{Ezawa:1983gp,Bais:1984xb,Lee:1984ica,Xu:1988zj,Mann:2005gk,Cotaescu:2005jg}. 

The simplest topological solution in the GPS model is a magnetic monopole which arises from the generalization to more dimensions of the Taub-NUT solitonic solution \cite{Newman:1963yy,Hawking:1976jb} and is described by the following metric

\begin{equation}  
ds^2= -dt^2 + V (dx^5 +4m(1- \cos \theta) d \phi)^2 +\frac{1}{V} (dr^2 +r^2 d\theta^2 +r^2\sin^2 \theta d \phi^2),
\end{equation}
where $V^{-1} = 1 + \frac{4m}{r}$ and $r, \theta, \phi$ the spherical coordinates.  At large distances the gauge field from the GPS solution behaves as a Dirac monopole \cite{Dirac:1948um}

\begin{equation}
A_{\phi} \sim \frac{4m(1- \cos \theta)}{r \sin \theta}
\end{equation}
whose inertial mass $M$ is,

\begin{equation}
M \sim \frac{M_{Pl}}{4 \sqrt{\alpha}},
\label{Mass}
\end{equation}
where $M_{Pl}$ is  Planck's mass and $\alpha$ the fine structure constant. Note the non-perturbative character of this result in the inverse dependence on the structure constant. 

These KK monopoles have a characteristic, which distinguishes them from all other types of monopoles, that makes them interesting and is the leitmotiv of the present investigation. In ordinary Yang-Mills theories one does not find solitons which are stable magnetic dipoles because the superposition of a monopole and an anti-monopole has the topology of the vacuum and therefore nothing prevents them from annihilating. The situation is quite different in KK theories.  The vacuum solution belongs to a different topological sector than the monopole-anti-monopole solution  \cite{Gibbons:1979xm}. Since one cannot evolve smoothly from one type of solution to the other the monopole-anti-monopole pair will not annihilate classically. In the GPS scenario, KK dipoles can be constructed  by the Kerr-Schwarzschild metric, which are regular solutions in euclidean $3 +1$ dimensions, by compactifying the fourth dimension \cite{Gross:1983hb}. This provides a very interesting topological state which has no magnetic charge and behaves like a magnetic dipole  at large distances

\begin{equation}
A_\phi \sim \frac{4 M_d d \sin\theta}{r^2},
\end{equation}
\begin{figure}[htb]
\begin{center}
\includegraphics[scale= 0.8]{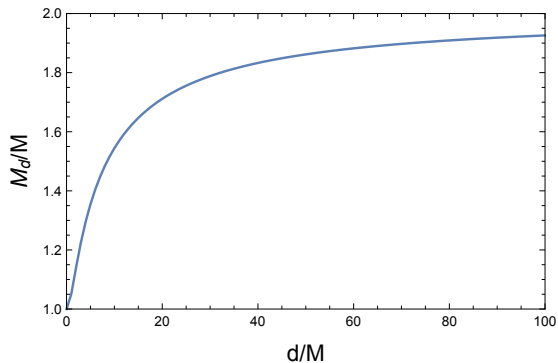} 
\end{center}
\caption{Monopolium mass $M_d/M$  as a function of the distance between the poles $d/M$.}
\label{Monopolium}
\end{figure}
where $M_d$ is the dipole mass and $d$ measures the distance between the poles.  For large $d$ this solution coincides with the product ansatz of the monopole-antimonopole solution. 

The mass of this dipole state is determined by the following equation,

\begin{equation}
M_d= 2  \left(M-\frac{M_d^2}{\sqrt{4M_d^2+d^2}}\right),
\label{monopoliummass}
\end{equation}
where we use the Planckian System of untits $\hbar= c = G = 1$. As can be seen in Fig. \ref{Monopolium} its mass ranges naturally from twice the monopole mass for large $d$ when the poles are far away from each other, to the monopole mass when the poles are strongly coupled  on top of each other. Thus a pair of monopole-anti-monopole if they get close enough to feel a strong attraction will make a neutral dipole state which we call monopolium. For  $d\sim 0$ this state is electromagnetically inert and will only interact gravitationally. Monopolium is classically stable and therefore may only decay quantum mechanically in a tunnel type process being therefore very long lived, i.e. almost stable. The model possesses other topological states of higher genus which at present are of no interest for our discussion. This study can be generalized to higher dimensions \cite{Ezawa:1983gp,Bais:1984xb,Lee:1984ica,Xu:1988zj,Mann:2005gk,Cotaescu:2005jg}. 

In the GPS model the inertial masses of monopole and dipole are different from the gravitational mass which is zero in a peculiar realization of the equivalence principle \cite{Gross:1983hb}. These strange gravitational properties of the Kaluza-Klein theory are connected with the existence of a pseudo-Goldstone massless scalar field, a dreaded dilaton,  as can be seen from the equivalent low energy theory (LET) \cite{Gross:1983hb}

\begin{equation}
{\mathcal S} = -\frac{1}{16 \pi G} \int{ d^4 x\left( R_4 + \frac{1}{6} g_{\mu \,\nu} \frac{\partial_\mu \sigma \, \partial_\nu \sigma}{\sigma^2} +\frac{1}{4} \sigma F_{\mu\, \nu}F^{\mu \, \nu}\right)}
\label{LET}
\end{equation}
where $\Phi \sim \log{\sigma}$ is the dilaton, which couples to gravity and electromagnetism.  Note that in the LET the masses of the other particles get multiplied by a conformal factor $\bar{\mu} = \mu \,\sigma^{-1/6}$. 

It has been hinted perturbatively, that in a full calculation quantum effects might provide the dilaton with a mass~\cite{Appelquist:1982zs}. Given the scales involved that mass would be at the level of the Planck mass and thus the dilaton field can be screened out of the theory.

 If we add a mass term to the above action $\frac{1}{2} M^2 \Phi^2$, the field equation for the scalar field becomes in the asymptotic limit ($r \rightarrow \infty$) of the spherical dipole metric \cite{Gross:1983hb},

\begin{equation}
\frac{\rho^4}{m^2}\left((1-2\,\rho)\frac{d^2 \Phi}{d \rho^2} -2\, \frac{d \Phi}{d \rho}\right) - \sqrt{(1-2\,\rho)} \, M ^2\ \Phi =0
\label{dilaton}
\end{equation}
where $\rho= \frac{m}{r}$, $m$ being a parameter of the metric and $r$ the spherical  radial coordinate. In the asymptotic limit $\rho \rightarrow 0$ and $ M $ is planckian. In the limit where the mass is effective, $r \rightarrow \infty$, i.e. $\rho \rightarrow 0$,  Eq.(\ref{dilaton}) leads to 
$\Phi \sim \exp{(- \frac{m M}{\rho})} \rightarrow 0$ and therefore $\sigma \rightarrow 1$ and thus $\bar{\mu} =\mu$. 

When the dilaton has no mass the equations of motion of a test particle of mass $\bar{\mu}$ lead to a cancellation between the newtonian force $\frac{1}{2} \bar{\mu} \partial_i (-g_{00}) U^0 U^0$ and  $\partial_i \bar{\mu}$, where $U^\mu = \frac{dx^\mu}{d\tau}$~\cite{Gross:1983hb}.  This cancellation is instrumental in the vanishing of the gravitational mass. However, once a dilaton mass is incorporated, the $\Phi$ field vanishes and $\bar{\mu}$ becomes approximately constant and there is no cancelation. Moreover since $M$ will be planckian the cancellation occurs for relatively small values of $r$.
Thus we expect that in any realistic KK theory where the dilaton has been screened out the equivalence principle will be realized in the conventional way and thus we assume that the inertial y gravitational masses are equal, noting that the inertial mass is what enters into our calculation. Moreover, although we will use the results of the GPS model, the only requirements for the anlysis that follows is that the monopole has a mass proportional to Planck's mass and that the dipole has a mass which varies from twice the monopole mass to a smaller mass when  the deformation of the dipole diminishes. The factors in Eq.(\ref{Mass}) or Eq.(\ref{monopoliummass}) are irrelevant given the huge numbers involved. Only the exponents count!

KK monopoles and monopolium have huge masses, larger than GUT  monopoles  \cite{Preskill:1984gd,Shnir:2005xx}, and therefore they have to be created at very early times and and their density will be strongly affected by inflation \cite{Guth:1980zm,Linde:1981mu}. Is there any chance of having some remnants at present? 

\section{Monopolium as a dark matter candidate} \label{dm}

Let us assume a cosmological model which expands isotropically in $3 + n$ spatial dimensions. A fluctuation of the geometry during this expansion causes $n$ dimensions to compactify. One may regard compactification as a phase transition and use the horizon distance as an order parameter \cite{Harvey:1984fe}. This results in the production of one KK monopole per horizon distance \cite{Zeldovich:1978wj,Preskill:1979zi}. If compactification occurs at $T \sim M $ then $d_H(T) \sim \frac{M_{PL}}{M^2} \sim \frac{16\,\alpha}{M_{Pl}}$ and in a radiation dominated expansion the ratio of the number density of KK monopoles, $n(T)$, to the number density of photons $n_{\gamma}(T)$ is \cite{Harvey:1984fe}

\begin{equation}
\frac{n(T)}{n_{\gamma}(T)} \sim \left(\frac{M}{M_{PL}}\right )^3 \sim \left(\frac{1}{4 \sqrt{\alpha}}\right)^3.
\end{equation}

 The interactions at formation between monopole and anti-monopole and with the charge in the plasma are dominated by the long range magnetic coupling.  Monopole and anti-monopole after diffusing through the plasma will bind instead of annihilate \cite{Preskill:1979zi}. The pair builds a highly excited monopolium of mass approximately $M_d \sim 2 M$ and large inter-pole distance $d$. The strong magnetic coupling will make monopole and anti-monopole cascade down to the ground state whose mass is $M_d \sim M$ and whose separation distance $d \sim 0$ (see Fig. \ref{Monopolium}). This ground state is electromagnetically inert, i.e. zero magnetic charge and zero magnetic dipole moment. It only is subject to gravitational interaction. Since monopole and anti-monopole are produced in equal numbers the number density of monopolium initially is still $n(M)$. This is a huge number, $n(M) \sim \left(\frac{1}{4 \sqrt{\alpha}}\right)^3 n_{\gamma}(M) \sim  \;10^{107} \,m^{-3}$. Since monopolium is basically stable its density decrease via the simplified  Boltzman's equation without annihilation terms,

\begin{equation}
\frac{n(T)}{T^3} = \frac{n(T')}{T^{' 3}}\; \mbox{  for }  \; T' < T,
\label{evolution}
\end{equation}
leading to a number density incompatible with standard cosmology. Thus we are lead to inflation.

\begin{figure}[htb]
\begin{center}
\includegraphics[scale= 0.8]{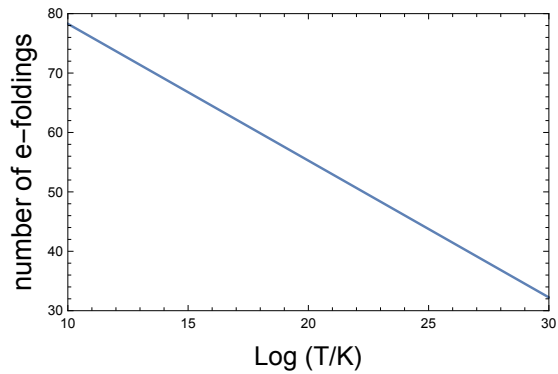} 
\end{center}
\caption{Number of e-foldings as a function of the decimal Log of the reheating temperature in Kelvin.}
\label{reheating}
\end{figure}

Before inflation at $T \sim M$ the density is

\begin{equation}
n(M) \sim 10^{107}\, m^{-3}.
\label{densitypreinflation}
\end{equation}
Assuming for the as an approximation that  all matter is made of monopolium we get,

\begin{equation}
n(2.7 K) = \frac{\rho_{matter}(2.7 K)}{M} \sim \;10^{-20}\, m^{-3}, 
\label{densitytoday}
\end{equation}
where $\rho_{matter}(2.7)$ is the matter density of the universe now. Let us assume that after inflation the reheating temperature of the universe is $T_{BB}$, then at the reheating period we get by Eq.(\ref{evolution}) a monopolium density

\begin{equation}
n(T_{BB})  \sim n(2.7 K) \left(\frac{T_{BB}}{2.7 K}\right)^3 .
\label{densityreheating}
\end{equation}
Thus from Eq.(\ref{densitypreinflation}) and  Eq.(\ref{densityreheating}) a direct relation can be established between the reheating temperature and the number of  e-foldings required from inflation as shown in Fig. \ref{reheating}. For example for a reheating temperature of $10^{27}K$ the number of required e-foldings is $37$. With these properties just discussed, monopolium in its ground state, could be an important candidate for constituent of dark matter or in any case an exotic primordial constituent of the universe.

\begin{figure}[htb]
\begin{center}
\includegraphics[scale= 0.85]{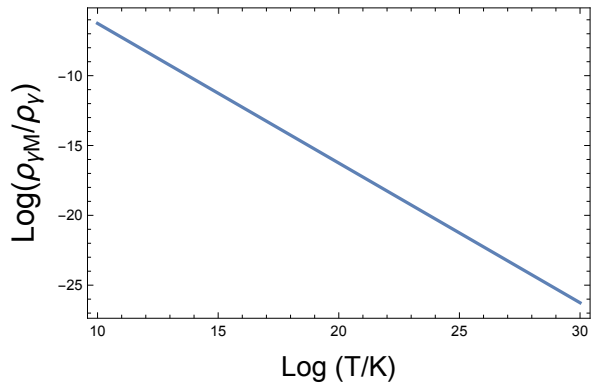} 
\end{center}
\caption{Decimal Log of the ratio of the energy density of the photons radiated from monopolium to the conventional blackbody photon radiation as a function on the decimal Log of the reheating temperature in Kelvin.}
\label{photons}
\end{figure}

Can the de-excitation of monopolium from the excited state at which it is formed to the ground state by emitting photons influence  the present microwave background? The time scale for de-excitation is large compared with the inflation period and therefore this process occurs during the reheating period. Assume the most favorable scenario that the excess mass from $2 M$ to $M$ is emitted as photons. In Fig. \ref{photons} we show the ratio of the energy density of the radiated photons during de-excitation $\rho_{\gamma M}(T) = M n (T)$ given by,

\begin{equation}
 \rho_{\gamma M}(T) = 0.26 \; 10^{-1} \left( \frac{T}{K}\right)^3 \frac{GeV}{m^3},
 \end{equation}
to the true photon density determined by

\begin{equation}
\rho_{\gamma}(T) = 0.47\, 10^{-5} \left( \frac{T}{K}\right)^4 \frac{GeV}{m^3}.
\end{equation}
As shown in the figure the emitted photons do not contribute significantly to reheating since they are a small fraction of  the photons associated  to conventional blackbody radiation.

Our analysis leads to two possible scenarios depending on the size of the  inflationary process and subsequent reheating period. If the inflationary process is around 37 e-foldings  monopolium could be a major constituent of dark matter. However, if  inflation is more dramatic, then the density of monopolium in the Universe will be reduced considerably, but monopolium might still form clouds of primordial non interacting matter. The first scenario is extremely exciting because it opens a very accessible door to extra dimensions. The second scenario is more modest but also worth pursuing.

\section{Signals of monopolium matter} \label{signals}

Since monopolium is made of magnetically charged particles one might wonder how monopolium reacts not in empty space but in the midst of the magnetic fields that crowd the universe. Does its behavior give rise to observable signals that discard monopolium as a dark matter particle? 

Monopolium is a monopole-anti-monopole bound state and in its ground state does not have a permanent dipole moment. However, in the vicinity of a magnetic field it can get an induced magnetic dipole moment through its response to the  external magnetic field. 

\begin{equation}
\overrightarrow{\cal{M}} = \alpha_M \vec{B} = 2 g \, \overrightarrow{\Delta d},
\label{effect}
\end{equation}
where  $\overrightarrow{\cal M}$  is the magnetic dipole induced in the system, $ \vec{B} $ the magnetic field,  $\alpha_M$ defines the magnetic (paramagnetic) susceptibility and $\Delta d =|\overrightarrow{\Delta d}|$ is the induced variation of the pole distance and therefore a measure of the stiffness of monopolium.

When clouds of monopolium matter traverse magnetic fields the induced magnetic moment will transform  the space fluctuations of the magnetic field into a time dependence that will generate electromagnetic radiation.  
Assume for simplicity that the time dependence is harmonic,

\begin{equation}
{\cal{M}}={ \cal{M}}_0 \cos{\omega t}.
\end{equation}
In this case monopolium matter will radiate  with a power proportional to the size of the induced magnetic moment \cite{Jackson:1998nia}
\begin{equation}
W = \frac{ k^4 {\cal{M}}_0^2}{3}
\label{power}
\end{equation}
where $k$ is the modulus of the wave vector $k=|\vec{k}|=\frac{\omega}{c}= \frac{2 \pi}{\lambda}$.

Thus as a cloud of monopolium matter approaches a magnetic field the region close to it will emit radiation while the one far away will not. That behavior will distinguish monopolium matter from other types of inert matter. In general the emission will not be harmonic but any emission can be formulated as a combination of harmonics. The crucial parameter for observability is the magnetic moment amplitude ${\cal M}_0$.

For a spring of constant $\kappa$ between two charged magnetic poles the polarizability is

\begin{equation}
\alpha_M = \frac{g^2}{\kappa}.
\end{equation}
as can be shown classically and quantum mechanically.

We use the Born-Oppenheimer approximation to find the spring constant of monopolium at short distances. From  Eq.(\ref{monopoliummass}) the $d << M$ limit leads to

\begin{equation}
\frac{M_d}{M} \sim 1+ \frac{1}{16} \frac{d^2}{M^2}.
\end{equation}
Transforming $d$ to length dimensions becomes

\begin{equation}
M_d = M + \frac{1}{16} \frac{d^2}{M  G^2}
\end{equation}
where the masses are measured in $GeV$, $d$ in $fm$ and we use Newton's constant as $G=1.32 \; 10^{-39}\; \frac{fm}{GeV}$. 
The spring constant  becomes  therefore $\kappa= \frac{1}{8 M G^2}$, thus  $\alpha_M = 8 M g^2 G^2 $ and 

\begin{equation}
{\cal M}_0 = 8 M g^2 G^2 B,
\end{equation}
where $B$ is measured in $GeV^2$.

Using the Dirac's quantization condition as $\frac{e g}{\hbar c} =\frac{1}{2}$ and the fine structure constant $\alpha = \frac{e^2}{\hbar c}$ we get
~\cite{Jackson:1998nia} 

\begin{equation}
\alpha_M = 8 \left(\frac{\hbar c}{2e}\right)^2 M G^2 \sim 54\; M G^2 \; fm^3,
\label{alphaM}
\end{equation}
thus

\begin{equation}
{\cal M}_0 \sim 1.4 \; 10^3 M  G^2 B\; fm.
\label{moment}
\end{equation}
To summarize, monopolium matter which is electromagnetically inert in the interstellar vacuum  in the presence of time dependent magnetic fields emits radiation as determined by Eqs.(\ref{power}) and (\ref{moment}). This radiation is a signal that differentiates monopolium clouds from other exotic matter clouds. Is this signal big enough to be detected?  If affirmative it would require the planning of experiments to confirm their existence. If not observable, monopolium  could be a possible dark matter constituent. The aim of the next section is to estimate the size of that emission.

\section{Astrophysical scenario} \label{astro}

\begin{figure}[htb]
\begin{center}
\vspace{-1.0cm}
\includegraphics[scale= 0.4,angle=90]{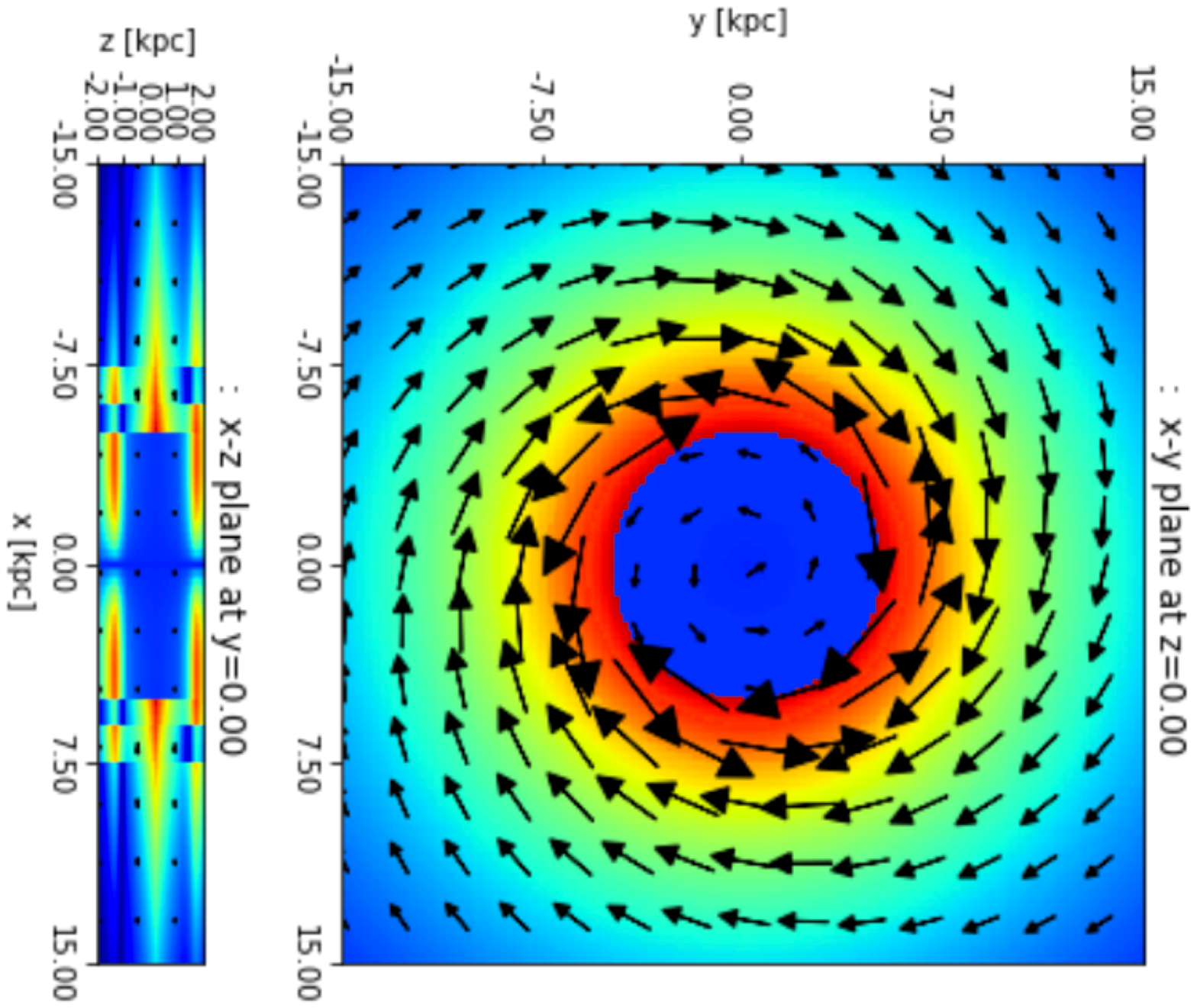} 
\end{center}
\vspace{-1.5cm}


\caption{Schematic picture of the magnetic field in our model galaxy. $B$ is measured in microgauss ($\mu$G) and $x, y ,z$ in kiloparsecs (Kpc).}
\label{galaxy}
\end{figure}

In order to answer the question posed in the previous section we  analyze the most favorable astrophysical scenario by means of a simple model.  Imagine a cloud of monopolium matter moving towards the center of a galaxy. We use a very simple model for the magnetic core of certain type of galaxies. We assume a basically flat galaxy, circular in shape, and which contains two regions, the external region with radius $R$ such that $R_G> R > R_C$, where the magnetic field is oriented in a clockwise direction, and the internal region with radius $R$ such that $R < R_C$, where  the magnetic field points in an anti-clockwise direction as shown in Fig. \ref{galaxy}. In the transition region the field is maximal and its direction rapidly flipping.

\begin{figure}[htb]
\begin{center}
\includegraphics[scale= 0.85]{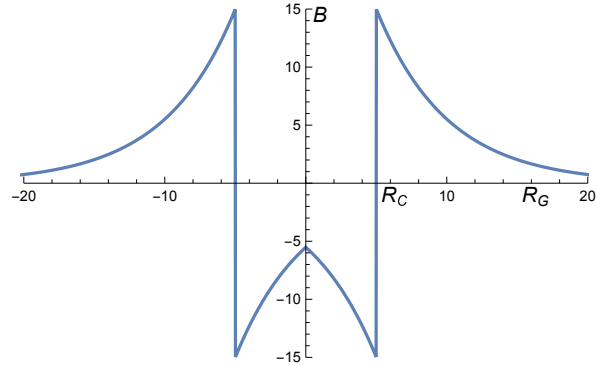} \end{center}
\caption{One dimensional model of the magnetic field of a galaxy used in the calculation as a function of distance.}
\label{Bfield}
\end{figure}

This type of models have been studied in the literature \cite{Jaffe:2019iuk} but we simplify them further for our calculation. Let us assume that a monopolium cloud approaches the galaxy from the right and that the monopolium cloud is small enough compared with the size of the galaxy ($ \sim 1\, \mbox{Kpc}$) that we can treat it as a point like object with mass $M_C$. Thus we reduce the effective B field for the calculation  to the one  shown in Fig. \ref{Bfield}. 

\begin{figure}[htb]
\begin{center}
\includegraphics[scale= 0.85]{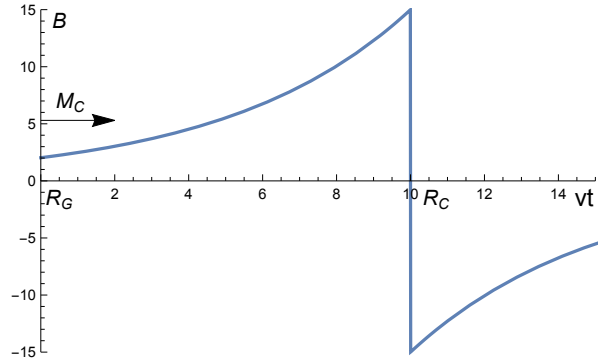} 
\end{center}
\caption{Representation of the time evolution of the magnetic field seen by the monopolium cloud.}
\label{monogalaxy}
\end{figure}

We can formulate mathematically the magnetic field by

\begin{eqnarray}
B(R)  = &\;\,B_0 \;\tanh{\left(\frac{R-R_C}{R_1}\right)} \exp{\left(-\left|\frac{R-R_C}{R_0}\right|\right)}  \hspace{1.5cm}&{\mbox  for} \; R > 0, 
\nonumber\\
  = &- B_0 \;\tanh{\left(\frac{R+R_C}{R_1}\right)} \exp{\left(-\left|\frac{R+R_C}{R_0}\right|\right)} \hspace{1.5cm}&{\mbox  for} \; R < 0 ,  \nonumber \\
  & 
 \end{eqnarray}
where $B_0 \sim 15\, \mu G$ the largest possible field found in galaxies, 
$, R_0 = 5 \,\mbox{Kpc}, R_G \sim 15\, \mbox{Kpc}$, $R_C \sim\, 5\, \mbox{Kpc}$ and $R_1 <<  R_0$. We have chosen  $R_1 = 1.0 \,\mbox{pc}$ to plot the figure. Let the monopolium cloud move from the right towards the center  with a velocity $v$. The moving cloud feels a time dependent magnetic field  $B(t) \sim B(R= R_G - v t)$, where $R_G$ is the point where we start measuring the time as can be seen in Fig. \ref{monogalaxy}. The most interesting region is close to $R_C$ as we shall see.

The power emitted by a non harmonic force is \cite{Mirmoosa:2020pft}
 
 \begin{equation}
 W(t) \sim {\cal M}(t) \frac{d^4 {\cal M}(t)}{dt^4} \sim \alpha_M^2 B(t) \frac{d^4 B(t)}{dt^4}.
 \end{equation}
 where ${\cal M}(t) = \alpha_M B(t)$, where recall that $\alpha_M$ is the magnetic polarizability of monopolium, Eq.(\ref{alphaM}). The power can be written as
 
 \begin{equation}
 W(t) \sim \alpha_M^2 B_o^2 \gamma^4 b(x) \frac{d^4 b(x)}{dx^4}
 \end{equation}
 where $x = \frac{R_G-R_C-v t}{R_1}$, $\gamma = \frac{v}{R_1}$, and 
 
 \begin{eqnarray}
 b(x)& = & \tanh{(x)} \exp{(-\,a\, x)} \;\;  \mbox{ if } \hspace{0.5cm} 0< x< \frac{R_G-R_c}{R_1} \nonumber \\
  &=& \tanh{(x)} \exp{(+\,a\, x)}\;\;  \mbox{ if } \;\;  -\frac{R_c}{R_1}< x< 0 \nonumber \\
  & & 
  \end{eqnarray}
being $a= \frac{R_1}{R_0}$.

\begin{figure}[htb]
\begin{center}
\includegraphics[scale= 0.72]{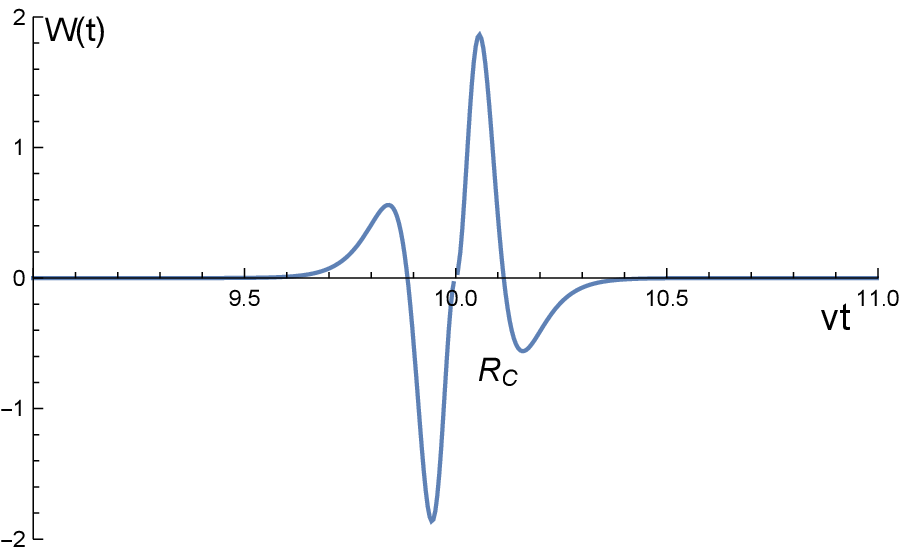} \hspace{1cm}\includegraphics[scale= 0.72]{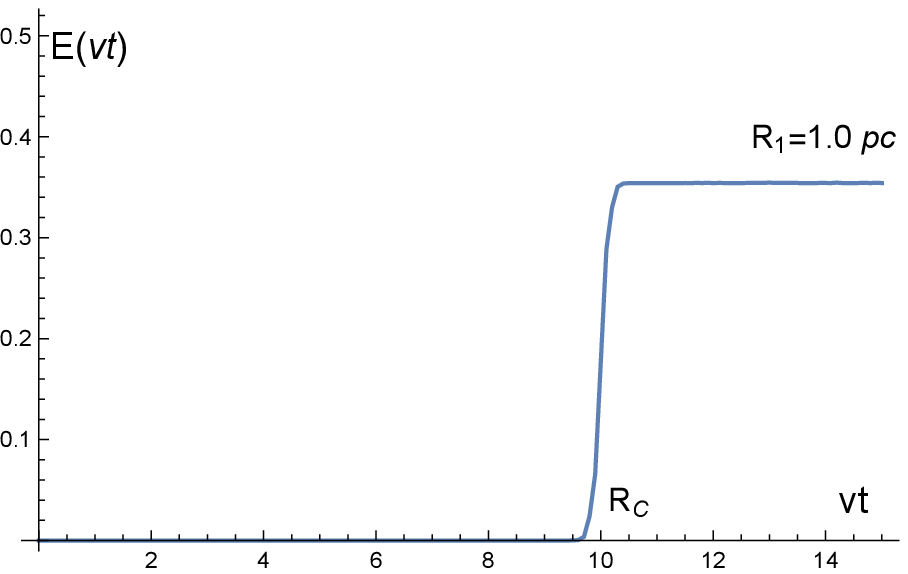} 
\end{center}
\caption{a) Emitted power taking into account the direction of the field. b) The structure of the emitted energy as a function of the penetration distance.}
\label{wave}
\end{figure}

The emission of power taking into account the magnetic field direction is shown in Fig.\ref{wave}a which is the emitted wave signal. The total energy emitted in the process at the point $R_G -v t$ is

\begin{equation}
E(vt) \sim  \int_0^{vt} d v t' W(v t') 
\end{equation}

In Fig. \ref{wave}b we show the energy emitted as the cloud penetrates the galaxy noticing that most of the emission occurs precisely where the power wave is formed around $R_C$ and approximately during the interval  $\sim [R_C - R_1, R_C + R_1]$.

If we integrate over the whole interval from $R_G$ to the origin we get

\begin{equation}
E (R_G) \sim \alpha_M^2 B_0^2 \gamma^3 { \cal I} (R_G, R_C, R_0, R_1)
\label{energy}
\end{equation}
where

\begin{equation}
{ \cal I} (R_G, R_C, R_0, R_1) =  \int_{- \frac{R_C}{R_1}}^{\frac{R_G-R_C }{R_1} } dx\, b(x) \frac{d^4b(x)}{dx^4}.
\end{equation}
Note that ${\cal I}$ is a numerical integral which dependes on the shape of the magnetic field and the structure of the galaxy as determined by the galaxy parameters $R_G, R_C, R_0, R_1$ but does not depend on the velocity. The only velocity dependence is in the factor $\gamma$ in front of Eq.(\ref{energy}). $R_1$ is the less known parameter but it is smaller than the rest of size parameters.  We use for the other size parameters realistic values as shown in the figures $B_0= \,15 \mu \mbox{G}$, which is the value of the highest possible galactic magnetic fields observed in nature, $R_G \sim 15 \,\mbox{Kpc}$, $R_C = 5\, \mbox{Kpc}$, $R_0 \sim 5 \,\mbox{Kpc} $ and for $R_1 \sim 1 \, \mbox{pc}$. This small value of $R_1$ makes the transition from one field direction to the other very abrupt. 

\begin{figure}[htb]
\begin{center}
\includegraphics[scale= 0.8]{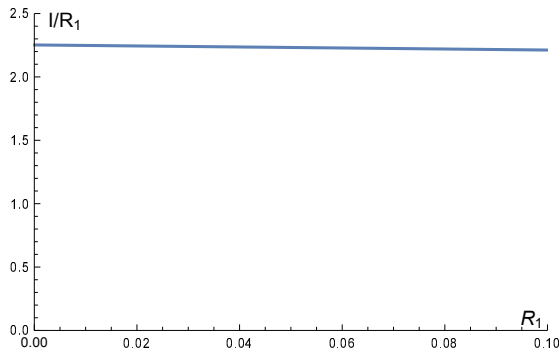} 
\end{center}
\caption{We plot the structure of the dependence of ${\cal I} (R_G, R_C, R_0, R_1) / R_1$ as a function of $R_1$ for small values of this parameter}
\label{R1I}
\end{figure}

Studying the dependence of the Integral on $R_1$ for small values of $R_1$ we find that it goes to a good approximation like a constant times $R_1$ as shown in Fig \ref{R1I}, i.e. ${\cal I} (R_G, R_C, R_0 , R_1) / R_1 \sim \mbox{constant} $. For the chosen values of the parameters the constant factor is approximately $2$.

The emitted energy when the monopolium cloud passes across $R_C$,

\begin{equation}
E \sim 2 \alpha_M^2 B_0^2 \gamma^3 \left(\frac{R_1}{\mbox{Kpc}}\right) n(T) V_{cloud}
\end{equation}
where $n(T)$ is the monopolium density which we parametrize by a temperature and its determined by the relation

\begin{equation}
n(T) = n(2.7 K) \left(\frac{T}{2.7}\right)^3
\end{equation}
where $n(2.7)$ is the monopolium density today as shown in Eq.(\ref{densitytoday}), and $V_{cloud}$ is the volume of the monopolium cloud. Assuming a spherical cloud we get for the energy emitted when the cloud passes across the region $R_C$

\begin{equation}
E = \frac{8\pi}{9}\left(\frac{1}{0.197}\right)\alpha_M^2 B_0^2\left(\frac{v}{c}\right)^3 n(2.7)\left(\frac{T}{2.7}\right)^3 \left(\frac{R_{cloud}}{R_1}\right)^3\left( \frac{R_1}{\mbox{Kpc}}\right) \mbox{GeV}
\end{equation}
where we have used the following units:  $\alpha_M$ is in fm$^3$, $B_0$ in GeV/fm, $n(2.7)$ in fm$^{-3}$ which lead to an energy in GeV. The duration in passing the region is governed by the cloud diameter, $\Delta T = 2 R_{cloud}/ v$.  Using a favorable case where $B_0 \sim15 \,\mu$G, $R_{cloud} \sim 1\,\mbox{Kpc}$, $R_1\sim \,1\mbox{pc}$, $v =0.01$ c, the emitted power is extremely small $P \sim 10^{-160}$ GeV/s $\sim 10^{-170} $ Joule/s at emission. The smallness is associated with the smallness of the polarizability $\alpha_m \sim 10^{-56}\, \mbox{fm}^3$ and also the small magnitude  of the galactic fields.  Thus monopolium matter associated with Kaluza-Klein monopoles is really dark and the fact that it is stable makes it very improbable to detect via its decay.

 \section{Concluding remarks}

 The existence of monopoles is a characteristic signature of Kaluza-Klein multidimensional theories \cite{Gross:1983hb,Sorkin:1983ns} based on fundamental developments in lower dimensional theories \cite{Newman:1963yy,Hawking:1976jb,Gibbons:1979xm}. The topology of these theories is extremely interesting, in particular the leitmotiv of my presentation is inspired by the existence of the dipole solution \cite{Gross:1983hb,Kerr:1963ud,Gibbons:1976ue}, which has been associated to a monopole-anti-monopole bound state. In KK theories this state is classically stable since it does not have the quantum numbers of the vacuum and therefore is long lived. This state has been named monopolium.  Its lowest energy state, i.e. $d\sim 0$ and $M_d \sim M$,  is electromagnetically inert in free space. We have studied its properties and therefore suggest to propose it as a dark matter constituent.

In order to get some consequences of this line of thought we have used the details of the GPS model which makes the presentation simpler by dealing with exact formulas. But the outcome would be the same accepting some very general features: the topology \cite{Newman:1963yy,Hawking:1976jb,Gibbons:1979xm,Kerr:1963ud,Gibbons:1976ue} ; the production mechanism by the compactification transition \cite{Zeldovich:1978wj,Preskill:1979zi,Harvey:1984fe} ; the naturalness of parameters in terms of  $M_{Pl}$ and $G$. The only formula used from GPS can simply substituted by $M_d \sim A + B d^2$, where $A\sim M_{Pl}$ and $B \sim (M_{Pl}G^2)^{-1}$ by dimensional arguments. The only difference with the present calculation would be numerical coefficients, which are almost irrelevant, since what matters here are exponents.

Due to its huge mass, $M \sim M_{Pl}$, KK monopoles can only be produced in large quantities via a compactification phase transition in the early Universe \cite{Harvey:1984fe}.  Their long range magnetic force will bind the KK poles form monopolium since monopoles and anti-monopoles are produced in equal quantitites. We  have shown that there is a good chance that monopolium survives in detectable amounts after a conventional inflationary period and that their photon emission arising by de-excitation from the initially highly excited state to the ground state  will not affect the cosmic microwave background. We have analyzed the observability of monopolium by using its electromagnetic properties induced in the presence of astronomical magnetic fields. We have seen that its magnetic polarizability is extremely small and thus detection is impossible even if appearing in  large monopolium clouds. At this point we cannot think of any more favorable scenario than this one. There are objects, like magnetars \cite{Kaspi:2017fwg}, with much higher magnetic fields, but these fields are of very small extension, at the level of kms, thus $V_{cloud}$ would be extremely small, and the emitted power even smaller than the one we have calculated. The other scenario we would like to study in the future is what happens with a cloud on monopolium in the proximity of a supermassive black holes in the center of the galaxies \cite{Ghez:2008ms,Gillessen:2008qv}, but at this point we do not see how this phenomenon would increase observability. 
Thus KK monopolium seems very difficult to detect and therefore an ideal candidate to be one of the constituents of dark matter. Probably  monopolium clouds will be only detectable  by gravitational waves arising from their interaction with black holes.

What really remains to investigate is the discovery of possible signals to detect KK monopoles or KK monopolium. In this work we have assumed their existence and have described some properties which lead to  monopolium as being a primordial dark matter constituent. But a proof, even a good argument of its existence is lacking. Its huge mass makes  production unfeasible in accelerators, the small magnitude of its electromagnetic effects and its small present day density make it difficult to detect by astronomical observatories. Thus, the aim next is to search for properties of KK monopoles or KK monopolium, which help in proposing experiments which might prove or disprove its existence.

\section*{Acknowledgement}
Discussions with Gabriela Barenboim, Huner Fanchiotti, Armando P\'erez, Sergio Palomares, Santiago Noguera, and Javier Vijande are acknowledged.  I would like to thank Carlos Garc\'{\i}a Canal and Marco Traini for a careful reading of the manuscript and valuable comments and suggestions, and Marco Traini also for sharing his profound knowledge on polarizable systems with me. Conversations with V. Vento-Bosch were inspiring to initiate this study. This work was supported in part  by the MICINN and UE Feder under contract FPA2016-77177-C2-1-P.

\end{document}